%
%
\def\ptitle{\tiny Study of Anharmonic Singular Potentials}
\font\tr=cmr12                          
\font\bf=cmbx12                         
\font\it=cmti12                         
\font\trbig=cmbx12 scaled 1500          
\font\tiny=cmr10                        
\output={\shipout\vbox{\makeheadline
                                      \ifnum\the\pageno>1 {\hrule}  \fi 
                                      {\pagebody}   
                                      \makefootline}
                   \advancepageno}

\headline{\noindent {\ifnum\the\pageno>1 
                                   {\tiny \ptitle\hfil
page~\the\pageno}\fi}}
\footline{}

\tr 
\def\bra{{\rm <}}    
\def\ket{{\rm >}}    
\def\nl{\hfil\break\noindent}  
\def\ni{\noindent}             
\def\np{\hfil\vfil\break}
\def\ppl#1{{\leftskip=10cm\noindent #1\smallskip}} 

\def\bra{{\rm <}} 
\def\ket{{\rm >}} 
\def\hi#1#2{$#1$\kern -2pt-#2} 
\def\hy#1#2{#1-\kern -2pt$#2$} 
\def\dbox#1{\hbox{\vrule 
\vbox{\hrule \vskip #1\hbox{\hskip #1\vbox{\hsize=#1}\hskip #1}\vskip #1 
\hrule}\vrule}} 
\def\qed{\hfill \dbox{0.05true in}} 

\baselineskip 15 true pt  
\parskip=0pt plus 5pt 
\parindent 0.25in
\hsize 6.0 true in 
\hoffset 0.25 true in 
\emergencystretch=0.6 in                 
\vfuzz 0.4 in                            
\hfuzz  0.4 in                           
\vglue 0.1true in
\mathsurround=2pt                        
\topskip=24pt                            
\newcount\zz  \zz=0  
\newcount\q   
\newcount\qq    \qq=0  

\def\pref#1#2#3#4#5{\frenchspacing \global \advance \q by 1     
    \edef#1{\the\q}{\ifnum \zz=1{\item{[{\the\q}]}{#2\ }{\bf #3}~{#4.}{~#5}\medskip} \fi}}

\def\prefm#1#2#3#4#5#6#7{\frenchspacing \global \advance \q by 1     
    \edef#1{\the\q}{\ifnum \zz=1{\item{[{\the\q}]}{#2,~}{\it #3,~}{#4~}{\bf#5},
{#6.}{~#7}\medskip} \fi}}

\def\bref #1#2#3#4#5{\frenchspacing \global \advance \q by 1     
    \edef#1{\the\q}
    {\ifnum \zz=1 { %
       \item{[{\the\q}]} 
       {#2}, {#3} {(#4).}{~#5}\medskip} \fi}}

\def\gref #1#2{\frenchspacing \global \advance \q by 1  
    \edef#1{\the\q}
    {\ifnum \zz=1 { %
       \item{$^{\the\q}$} 
       {#2.}\medskip} \fi}}

 \def\sref #1{~[#1]}

\def\references#1{\zz=#1
   \parskip=2pt plus 1pt   
   {\ifnum \zz=1 {\noindent \bf References \medskip} \fi} \q=\qq

\pref{\case}{K. M. Case, Phys. Rev.}{80}{(1950)\ 797}{}
\pref{\rms}{R. M. Spector, J. Math. Phys.}{8}{(1967)\ 2357}{}
\pref{\spec}{R. M. Spector, J. Math. Phys.}{5}{(1964)\ 1185}{}
\pref{\klaa}{J. R. Klauder, Acta Phys. Austriaca Suppl.}{11}{(1973)\ 341 }{}
\pref{\Klab}{J. R. Klauder, Phys. lett. B}{47}{ (1973)\ 523}{}
\pref{\sim}{B. Simon, J. Functional Analysis}{14}{(1973)\ 295}{} 
\pref{\deh}{B. DeFacio and C. L. Hammer J. Math. Phys.}{15}{(1974)\ 1071}{}
\pref{\detw}{L. C. Detwiler and J. R. Klauder, Phys. Rev. D\ }{11}{(1975)\ 1436}{}
\pref{\harr}{E. M. Harrell, Ann. Phys. (NY)}{105}{(1977)\ 379}{}
\pref{\eks}{H. Ezawa, J. R. Klauder, and L. A. Shepp, J. Math. Phys.}{16}{(1975)\ 783}{}
\pref{\klac}{J. R. Klauder, Science}{199}{(1978)\ 735}{}
\pref{\agua}{V. C. Aguilera-Navarro, G.A. Est\'evez, and R. Guardiola, J. Math. Phys.}{31}{99 (1990)}{}
\pref{\agub}{V. C. Aguilera-Navarro and R. Guardiola, J. Math. Phys.}{32}{2135 (1991)}{}
\pref{\aguc}{V. C. Aguilera-Navarro, F. M. Fern\'andez, R. Guardiola and J. Ros, J.Phys. A: Math. Gen}{25}{6379 (1992)}{}
\pref{\agud}{V. C. Aguilera-Navarro, A. L. Coelho and Nazakat Ullah, Phys. Rev. A}{49} {1477 (1994)}{}
\pref{\sol}{Solano-Torres, G. A. Est\'eves, F. M. Fern\'andez, and G. C. Groenenboom, J. Phys. A: Math. Gen.}{25}{3427 (1992)}{}
\pref{\mdel}{M. de Llano, Rev. Mex. Fis.}{27}{(1981)\ 243 }{}
\pref{\kola}{H. J. Korsch and H. Laurent, J. Phys. B: At. Mol. Phys.}{14}{4213 (1981)}{}
\pref{\zb}{M. Znojil, J. Phys. lett.}{101A}{(1984)\ 66}{}
\pref{\znoa}{M. Znojil, J. Math. Phys.}{30}{(1989)\ 23}{}
\pref{\fer}{F. M. Fermendez, Phys. Lett. A}{160}{(1991)\ 511 }{}
\pref{\znod}{M. Znojil, Phys. Lett. A}{169}{(1992)\ 415 }{}
\pref{\zd}{M. Znojil and P. G. L. Leach, J. Math. Phys.}{33}{(1992)\ 2785}{}
\pref{\znoc}{M. Znojil, J. Math. Phys.}{34}{(1993)\ 4914}{}
\pref{\flyn}{M. F. Flynn, R. Guardiola, and M. Znojil, Czech. J. Phys.}{41}{(1993)\ 1019}{}
\pref{\mill}{H. G. Miller, J. Math. Phys.}{35}{2229 (1994)}{}
\pref{\znoe}{M. Znojil and R. Roychoudhury, Czech. J. Phys.}{48}{(1998)\ 1}{}
\pref{\znoe}{M. Znojil, Phys. Lett. A}{255}{(1999)\ 1 }{}
\pref{\nag}{N. Nag and R. Roychoudhury, Czech. J. Phys.}{46}{(1996)\ 343}{} 
\pref{\esta}{E. S. Est\'evez-Bret\'on and G. A. Est\'evez-Bret\'on,  J. Math. Phys.}{34}{(1993)\ 437}{}
\pref{\kila}{J. Killingbeck, J. Phys. A: Math. Gen.}{13}{(1980)\ 49}{}
\pref{\kilb}{J. Killingbeck, J. Phys. A: Math. Gen.}{13}{(1980)\ L231 }{}
\pref{\kilc}{J. Killingbeck, J. Phys. B: Mol. Phys.} {15}{(1982)\ 829}{}
\pref{\kild}{J. Killingbeck, J. Phys. A: Math. Gen.}{14}{(1981)\ 1005}{}
\pref{\kile}{J. Killingbeck, G. Jolicard and A. Grosjean, J. Phys. A: Math. Gen.}{34}{L367 (2001)}{}
\pref{\ma}{O. Mustafa and M. Odeh, J. Phys. B }{32}{(1999) 3055}{}
\pref{\mb}{O. Mustafa and M. Odeh, J. Phys. A }{33}{(2000) 5207}{}
\pref{\jski}{J. Skibi\'nski, e-print quant-ph/0007059}{}{}{}
\pref{\hnas}{R. Hall and N. Saad, Can. J. Phys.} {73}{(1995)\ 493}{}
\pref{\hala}{R. Hall, N. Saad and A. von Keviczky, J. Math. Phys.\ }{43}{(2002) 94} {}
\pref{\halb}{R. Hall, N. Saad and A. von Keviczky, J. Math. Phys. }{39}{(1998) 6345}{}
\pref{\halc}{R. Hall and N. Saad, J. Phys. A: Math. Gen. }{33}{(2000) 569}{}
\pref{\hald}{R. Hall, N. Saad and A. von Keviczky, J. Phys. A: Math. Gen. }{34}{(2001) 1169}{}
\pref{\hale}{N. Saad, R. Hall and A. von Keviczky, J. Math. Phys}{36}{(2003) 487}{}
\pref{\bb} {J. K. Bhattacharjee and S. Bhattacharyya, J. Phys. A: Math. Gen. }{36}{(2003) L223}{}
\pref{\roy} {Amlan K. Roy, Phys. Lett. A }{321}{(2004) 231}{}
\pref{\za} {M. Znojil, J. Math. Phys. }{31}{(1990) 108}{}
\pref{\zc} {M. Znojil, Phys. lett. A}{158}{(1991) 436 }{}
\pref{\zd} {M. Znojil, J. Phys. A: Math. Gen. }{15}{(1982) 2111 }{}
\pref{\pc} {Peace Chang and Chen-Shiung Hsue, Phys. Rev. A }{49}{(1994) 4448}{} 
\pref{\bgp} {E. Buend\'ia, F. J. G\'alvez and A. Puertas, J. Phys. A: Math. Gen. }{28} {(1995) 6731}{}
\pref{\ka} {R. S. Kaushal, Ann. Phys. (NY)}{206}{(1991) 90}{}
\pref{\kp} {R. S. Kaushal and D. Parashar, Phys. Lett. A }{170}{ (1992) 335 }{}
\pref{\gr} {R. Guardiola and J. Ros, J. Phys. A: Math. Gen. }{25} {(1992) 335}{}
\pref{\la} {M. Landtman, Phys. lett. A }{175} {(1993) 147}{}
\pref{\va} {Y. P. Varshni, Phys. Lett. A }{183} {(1993) 9}{}
\pref{\vb} {Y. P. Varshni, N. Nag, and R. Roychoudhury, Can. J. Phys. }{73}{(1995) 519}{}
\pref{\pm} {A. R. Plastino abd H. G. Miller, Phys. lett. A }{205}{(1995) 125}{}
\pref{\ak} {A. Khare and S. N. Behra, Pramana J. Phys. }{14}{(1980) 327}{}
\pref{\sz} {Shi-Hai Dong and Zhong-Qi Ma, J. Phys. A: Math. Gen. }{31} {(1998) 9855}{}
\pref{\var} {Y. B. varshni, Can. J. Phys. }{75}{(1997) 907-912}{}
\pref{\bar} {A. O. Barut, J. Math. Phys. }{21} {(1980) 568}{}
\bref{\br} {B. H. Bransden and C. J. Joachain}{Physics of Atoms and Molecules}{ (London: Longman) 1983}{}
\bref{\mr} {G. C. Mailtland, M. Rigby, E. B. Smith and W. A. Wakeham}{Intermolecular Forces}{ (Oxford: Oxford University Press) 1987}{}
\pref{\ze} {M. Znojil. J. Math. Phys. }{31}{(1990) 108}{}
\pref{\gotc} {B. G\"onu\"ul, O. \"Ozer, M. Ko\c{c}ak, D. Tutcu and Y. Can\c{c}elik, J. Phys. A: Math. Gen. }{34}{(2001) 8271-8279}{}
}

\references{0}    
\ppl{CUQM-106}\ppl{math-ph/0410039} 
\ppl{October 2004}\medskip 
\vskip 0.5 true in
\centerline{\bf\trbig Study of Anharmonic Singular Potentials}
\vskip 0.5true in

\centerline{\bf  Nasser Saad$\dag$, Richard L. Hall$^\dagger$, and Qutaibeh D. Katatbeh$^\ddagger$}
\medskip
{\leftskip=0pt plus 1fil
\rightskip=0pt plus 1fil\parfillskip=0pt
\obeylines
$\dag$ Department of Mathematics and Statistics,
University of Prince Edward Island, 
550 University Avenue, Charlottetown, 
PEI, Canada C1A 4P3.\par}
\medskip
{\leftskip=0pt plus 1fil
\rightskip=0pt plus 1fil\parfillskip=0pt
\obeylines
$^\dagger$ Department of Mathematics and Statistics, Concordia University,
1455 de Maisonneuve Boulevard West, Montr\'eal, 
Qu\'ebec, Canada H3G 1M8.\par}

\medskip
{\leftskip=0pt plus 1fil
\rightskip=0pt plus 1fil\parfillskip=0pt
\obeylines
$^\ddagger$ Department of Mathematics and Statistics,
Faculty of Science and Arts,
Jordan University of Science and Technology,
Irbid 22110, Jordan.\par}
\baselineskip 20 true pt

\vskip 0.2 true in
\centerline{\bf Abstract}\medskip
\noindent A simple and efficient variational method is introduced to accelerate the convergence of the eigenenergy computations for a Hamiltonian $H$ with singular potentials. Closed-form analytic expressions in $N$ dimensions are obtained for the matrix elements of $H$ with respect to the eigenfunctions of a soluble singular problem with two free parameters $A$ and $B$. The matrix eigenvalues are then optimized with respect to $A$ and $B$ for a given $N$. Applications, convergence rates, and comparisons with earlier work are discussed in detail.
\medskip
\nl{\bf Keywords:}~~Singular potentials, Spiked Harmonic Oscillator, Anharmonic Potentials, Variational Method.
\nl{\bf PACS } 03.65.Ge
\np
\baselineskip 20 true pt

\ni{\bf 1. Introduction }
\medskip
\noindent Sources of ongoing interest\sref{\case-\gotc} in the study of singular potentials are at least threefold: (i) regular Rayleigh-Schr\"odinger perturbation theory can fail badly for such potentials, and (ii) in physics, it is quite common to encounter phenomenological potentials that are strongly singular at the origin, and (iii) because of the intrinsically interesting mathematical problems that arise in their study. A specific family of singular quantum Hamiltonians that has found widespread application in many areas of atomic, molecular, nuclear physics, are the so-called anharmonic singular Hamiltonians given by
$$H=-{d^2\over dr^2}+r^2+\sum\limits_{n=0}^N {{\lambda_n}\over{r^{\alpha_n}}},\eqno(1.1)$$
where $r\in(0,\infty)$ and $\alpha_n$ and $\lambda_n$ are positive real numbers.   A particular and important special subclass of (1.1) that has been a subject of intensive studies is the set of spiked harmonic oscillator Hamiltonians\sref{\harr-\spec}
$$H=-{d^2\over dr^2}+r^2+{\lambda\over r^\alpha},\quad \alpha>0,\lambda>0\eqno(1.2)$$
acting in the Hilbert space $L_2(0,\infty)$ with eigenfunctions $\psi \in L_2(0,\infty)$ of $H$ satisfying the Schr\"odinger equation
$
-\psi^{\prime\prime} +(r^2+\lambda r^{-\alpha})\psi = E\psi\hbox{ with }\psi(0) = 0.
$
The function $\psi$ is an eigenfunction corresponding to the eigenvalue $E$ and the condition $\psi(0) = 0$ is the {\it Dirichlet boundary condition} we impose. The coupling parameter $\lambda$ determines the strength of the perturbative potential and the positive constant $\alpha$ represents the type of singularity at the origin. Thanks to the pioneering work of Detwiler and Klauder \sref{\detw} and of Harrell\sref{\harr}, remarkable progress has been made over the past three decades in the field of spectral calculations for the spiked harmonic oscillator Hamiltonian (1.2). Other interesting subclasses of (1.1) which have been used in atomic, molecular and nuclear physics are the anharmonic singular Hamiltonians \sref{\za-\var}
$$H=-{d^2\over dr^2}+ar^2+b r^{-4}+c r^{-6},\quad a>0,\ b>0,\ c>0\eqno(1.3)$$
and the positive-parameter singular even-power Hamiltonians\sref{\bar-\gotc}
$$H=-{d^2\over dr^2}+a_1r^2+{a_2\over r^2}+{a_3\over r^4}+{a_4\over r^6}.\eqno(1.4)$$
   Several numerical and analytical (both variational and perturbative) techniques are avaliable in the literature for the exact and approximate eigenvalue calculations for these families of Hamiltonians. Some of these techniques were devoted to particular classes of singular Hamiltonians and some others were restricted to specific values of the potential parameters. It is of great interest to have a successful method valid to study all of these classes without the need for major adjustment as one goes from class to class. The purpose of this article is to provide such technique and to present a rigorous variational approach for the accurate calculation of the energy levels of the singular Hamiltonians (1.1). Our method is very simple and yet accurate enough to determine the entire spectrum in arbitrary dimensions of anharmonic singular Hamiltonians (1.1). It can be viewed as an extension of the earlier variational approach to the study of the spiked Harmonic oscillator potentials\sref{\hale}. The principal ingrediants are (a) the use of a special basis comprising the exact solutions of a singular problem, (b) the determination of exact analytical expressions for the matrix elements with respect to this basis, and (c) the retention of two parameters from the basis problem that can be used as additional variational parameters.  These optimizations are made possible because of the progress in the establishment and simplification of closed-form analytic expressions for the matrix elements.

The article is organized as follows. In Section 2, we introduce our variational technique: the method is discussed in a general setting with no reference to particular application. In Section 3, we introduce the Gol'dman and Krivchenkov Hamiltonian as a solvable model. Thereafter, we use its eigensolutions to compute the matrix elements for more general singular operators $r^{-\alpha},\alpha>0$. Closed analytical expressions in terms of single finite sums are obtained for the matrix elements of the power-law potentials $r^{q}, q=2,4,6,\dots$. In Section 4, the applications to singular potentials are discussed. Special attention is paid to spiked harmonic oscillator (1.2) where we compared our results with other techniques avaliable in the literature. Thereafter, we study various higher-order anharmonic singular potentials, such as (1.3) and (1.4), using the approach discussed in Section 2. The convergence problem of the variational approach is studied in some detail. Comparisons with different methods for special classes of anharmonic singular potentials are also studied.
\medskip
\ni{\bf 2. Methodology}
\medskip
\noindent In this section, we develop a detailed variational method for studying the family of Hamiltonians (1.1). The variational function is taken as a linear combination of orthonormal functions of an exactly solvable model which itself has a singular potential. Let $\{E_n(\Omega),\psi_n^\Omega\}$ be the eigenvalues and the eigenfunctions of an exactly solvable Hamiltonian $H^\Omega=-\Delta+v^{\Omega}(r)$ acting on separable Hilbert space $L_2(D,d\mu)$ such that
$$\cases{H^\Omega\psi_n^\Omega=E_n(\Omega)\psi_n^\Omega,& \cr
\cr
	\parallel \psi_n^\Omega\parallel_D^2=\bra \psi_n^\Omega|\psi_n^\Omega\ket =\int\limits_D |\psi_n^\Omega(r)|^2d\mu(r)=1. & \cr}\eqno(2.1)
$$
Here $v^\Omega(r)$ is taken as a function of $r$ and depends on the parameters in the set $\Omega$.  Let $H^\Lambda=-\Delta+V^\Lambda(r)$ be the quantum Hamiltonian under investigation, where  $\Lambda$ is a fixed set of parameters, and we let $\epsilon(\Lambda)$ be the associated exact eigenvalues of $H^\Lambda$. By writing the Hamiltonian $H^\Lambda$ in the extended form 
$$H^\Lambda=H^\Omega+V^\Lambda(r)-v^\Omega(r),\eqno(2.2)$$
we have that if the integrals
$$\bra\psi_n^\Omega|V^\Lambda(\cdot)|\psi_n^\Omega\ket=\int\limits_D \overline{\psi_n^\Omega}V^\Lambda\psi_n^\Omega d\mu \hbox{\quad are finite},\eqno(2.3)$$
then the matrix elements of the Hamiltonian $H^\Lambda$ can be written, for $m,n=0,1,2,\dots$ as
$$H^\Lambda_{mn}\equiv\bra\psi_m^\Omega|H|\psi_n^\Omega\ket =E_{mn}(\Omega)\delta_{mn}+\bra \psi_m^\Omega|V^\Lambda(\cdot)|\psi_n^\Omega\ket-\bra \psi_m^\Omega|v^\Omega(\cdot)|\psi_n^\Omega\ket,\eqno(2.4)$$
or in compactified form as
$$H^\Lambda_{mn} =\bra \psi_n^\Omega|-\Delta|\psi_n^\Omega\ket+\bra \psi_n^\Omega|V^\Lambda(\cdot)|\psi_n^\Omega\ket.\eqno(2.5)$$

\noindent Our variational technique is based on forming trial wave functions from a linear combination of $D$-orthonormal functions $\psi_n^\Omega(r)$,$n=0,1,2,\dots, D-1$ 
$$\Psi(r)=\sum\limits_{n=0}^{D-1}c_n\psi_n^\Omega(r)\eqno(2.6)$$ 
The linear parameters $c_n$ that optimize the energy are determined by the following system of equations
$$\sum\limits_{n=0}^{D-1}(H^\Lambda_{mn}-\epsilon(\Lambda)\delta_{mn})c_n=0,\quad m=0,1,2,\dots,D-1,\eqno(2.7)$$
The necessary and sufficient condition for a nontrivial solution of (2.7) is the vanishing of the secular determinant
$$\hbox{det}|H^\Lambda_{mn}-\epsilon(\Lambda)\delta_{mn}|=0,\eqno(2.8)$$
The condition (2.8) yields upper bounds to the exact eigenvalues $\epsilon(\Lambda)$ by means of the inequality
$$\epsilon(\Lambda)\leq \min_\Omega\ \hbox{diag}\pmatrix{H^\Lambda_{00}&H^\Lambda_{01}&\dots&H^\Lambda_{0D-1}\cr
		    H^\Lambda_{10}&H^\Lambda_{11}&\dots&H^\Lambda_{1D-1}\cr
		    \dots&\dots&\dots&\dots\cr
		    H^\Lambda_{D-10}&H^\Lambda_{D-11}&\dots&H^\Lambda_{D-1D-1}},\eqno(2.9)$$
where $H^\Lambda_{mn}=\bra\psi_m^\Omega|H^\Lambda|\psi_n^\Omega\ket=H_{mn}^\Omega+V_{mn}^\Lambda(r)-v_{mn}^\Omega(r)$, for fixed $\Lambda$, are functions of the parameter set $\Omega$. Note that the equality in (2.9) holds if $V^\Lambda(r)= v^\Omega(r)$ $\mu$-a.e. in $r$. The computation of the right hand side of (2.9) requires a diagonalization of the matrix over the $D$-dimensional subspace spanned by orthonormal functions $\psi_n^\Omega$ followed subsequently by a minimization over the parameters $\Omega$. The advantages of this method are: (i) only a few matrix elements are needed to achieve accurate bounds to the eigenvalues; (ii) The exact eigenvalues are approached monotonically as $D$ is increased; (iii) the minimization over a set of parameters $\Omega$ accelerates the convergence of the energy bounds more rapidly than any standard minimization over a single variable; (iv) the optimization of (2.9) for $D\times D$ matrix gives upper bounds to the energy eigenvalues of the lowest $D$-states; (v) the diagonalization of such a matrix also produces the coefficients required for the corresponding eigenvectors determined variationally. 
\medskip
\ni{\bf 3. Exactly solvable model and associated matrix elements}
\medskip\noindent  The accuracy and the computational simplicity of the variational method depends greatly on the analytic structure of the wave functions that we use, in particular, their behaviour in the neighborhood of the singularity.  Many different forms of trial wave function have been explored in the literature to solve the spiked harmonic oscillator problem (1.2). The rate of convergence for a variational calculation depends on the ability of the basis functions used in the variational calculation to approximate the behavior of the exact wave function in the neighborhood of the singularity. Recently, Hall {\it et al} have pointed out the advantages of basing the variational analysis of singular potentials on an exact soluble model which itself has a singular potential term. They have suggested and used trial wave functions constructed by means of the superposition of the orthonormal functions of the exact solutions of  the Gol'dman and Krivchenkov Hamiltonian 
$$H_0=-{d^2\over dr^2}+Br^2+{A\over r^2}.\eqno(3.1)$$
It was shown that this orthonormal basis serves as an effective starting point for the variational analysis of the Hamiltonian (1.2). In this paper we use these solutions of $H_0$ to provide systematic variational solutions for the singular Hamiltonians (1.1). 
 The Gol'dman and Krivchenkov Hamiltonian (3.1) is one of the few tht admit exact analytical solutions. The Hamiltonian is the generalization of the familiar harmonic oscillator in 3-dimension $ -{d^2/dr^2}+Br^2+{l(l+1)/r^2}$ where the generalization lies in the parameter $A$ ranging over [$0,\infty)$ instead of only values determined by the angular momentum quantum numbers $l=0,1,2,\dots$. The background on  Gol'dman and Krivchenkov potential $V(r)=Br^2+Ar^{-2}$ relevant to the following discussion can be found in\sref{\hale}. In particular, the energy spectrum of the Schr\"odinger Hamiltonian $H_0$ is given, in terms of parameters $A$ and $B$, by
$${E}_n=2\beta(2n+\gamma),\quad n=0,1,2,\dots,\eqno(3.2)$$
in which $\beta=\sqrt{B}$ and $\gamma=1+\sqrt{A+{1\over 4}}$ and the normalized wavefunctions are
$$
\psi_n(r)=(-1)^n\sqrt{{2\beta^\gamma (\gamma)_n}\over n! \Gamma(\gamma)} r^{\gamma-{1\over 2}} e^{-{1\over 
2}\beta r^2}{}_1F_1(-n;\gamma;\beta r^2).\eqno(3.3)
$$ 
Here ${}_1F_1$ is the confluent hypergeometric function
$${}_1F_1(-n;b;z)=\sum\limits_{k=0}^n {{(-n)_kz^k}\over {(b)_kk!}},\quad\hbox{($n$-degree polynomial in $z$)}\eqno(3.4)$$
and the shifted factorial $(a)_n$ is defined by
$$(a)_0=1,\quad (a)_n=a(a+1)(a+2)\dots (a+n-1), \quad{\rm for}\ n = 1,2,3,\dots.\eqno(3.5)$$
may be expressed in terms of the Gamma function by $(a)_n={\Gamma(a+n)/ \Gamma(a),}$ when $a$ is not a negative integer $-m$, and, in these exceptional cases, $(-m)_n = 0$ if $n > m$ and otherwise $(-m)_n = (-1)^n m!/(m-n)!.$
 
An important observation regarding the solvable model $H_0$ is the existence of the $A$-term which has the dimensions of kinetic energy such as the term that appears in higher-dimensional systems. This observation allow us to extend (3.2) and (3.3) to the exact solutions of $N-$dimensional Gol'dman and Krivchenkov Hamiltonian, namely, 
$$-{d^2\over dr^2}+ {\Lambda(\Lambda+1)+A\over r^2} +Br^2, \quad (A\geq 0, B>0)\eqno(3.6)$$
where $\Lambda={1\over 2}(M-3)$, and $M=N+2l$. The exact solution of (3.6) can be easily found by replacing  $A$ in (3.2) and (3.3) with 
$$A\rightarrow \Lambda(\Lambda+1)+A.\eqno(3.7)$$
This particular observation can be extended to any  $3$-dimensional exact solvable quantum model. Indeed, if Schr\"odinger's equation can be solved for arbitrary angular momentum number $l$, then the extension to the $N$-dimensional case can be obtain by replacing $l$ with $\Lambda$. It should be also noted that $N$ and $l$ enter into the Hamiltonian (3.6) in the form of combination $N+2l$. Hence, the energy for a spherically-symmetric potential $V(r)$ are the same as long as $M$ is not altered.  We now summarize the exact eigenvalues of $N$-dimensional Schr\"odinger equation with the Gol'dman and Krivchenkov potential as
$$
E_{nl}^N=2\beta(2n+{\gamma}_N),\quad n,l=0,1,2,\dots,\eqno(3.8)
$$
where $\beta=\sqrt{B}$ and $\gamma_N=1+\sqrt{A+(\Lambda+{1\over 2})^2}$, while the exact eigenfunctions are given explicitly by
$$
\psi_{nl}^N(r)=(-1)^n\sqrt{{2\beta^{\gamma_N} (\gamma_N)_n}\over n! \Gamma(\gamma_N)} r^{\gamma_N-{1\over 2}} e^{-{1\over 
2}\beta r^2}{}_1F_1\bigg(\matrix{-n\cr
	\gamma_N\cr}\bigg|\beta r^2\bigg),\quad (n,l=0,1,2,\dots, N\geq 1).\eqno(3.9)
$$
In the next section all our results are forumlated in arbitrary dimension $N \geq 1.$
\medskip

\ni{\bf 3.1 Matrix elements of the singular operator $r^{-\alpha}$:}
\medskip\noindent  The effectiveness of the variational method relies on finding a basis that allows for easy calculation of the matrix elements of the given Hamiltonian. An important advantage of the orthonormal wavefunctions (3.3) is the existence of closed-form formulas for the singular potential integrals $\bra \psi_m|r^{-\alpha}|\psi_n\ket$. These closed form expressions are  achieved by means of the following identity:
For $m$ and $n$ non-negative integers and $2\gamma>\alpha$ 
$$\int\limits_0^\infty r^{2\gamma-\alpha-1} e^{-\beta r^2}
{}_1F_1\bigg(\matrix{-n\cr
	\gamma\cr}\bigg|\beta r^2\bigg){}_1F_1\bigg(\matrix{-m\cr
	\gamma\cr}\bigg|\beta r^2\bigg)dr={({\alpha\over 2})_n\Gamma(\gamma-{\alpha\over 2})\over2\beta^{\gamma-{\alpha\over 2}}(\gamma)_n} {}_3F_2\bigg(\matrix{-m,{\gamma-{\alpha\over 2}},{1-{\alpha\over 2}}\cr
	\gamma,1-{\alpha\over 2}-n\cr}\bigg|1\bigg)
\eqno(3.10)
$$
where the Clausen hypergeometric function ${}_3F_2$
is defined by the series representation
$${}_3F_2\bigg(\matrix{-m,a,b\cr
	c,d\cr}\bigg|1\bigg)=\sum\limits_{k=0}^m{(-m)_k(a)_k(b)_k\over (c)_k(d)_k\ k!},\quad (m-\hbox{degree polynomial)}.$$
The proof of this identity and some relevant integrals can be found in \sref{\hale}.
Thus the matrix elements $r_{mn}^{-\alpha}=\bra\psi_n|r^{-\alpha}|\psi_n\ket$ of the singular operator $r^{-\alpha}$ have the explicit forms
$$
r_{mn}^{-\alpha}=(-1)^{n+m}\beta^{{\alpha\over 2}}{{({\alpha\over 2})_n}\over
(\gamma)_n}{{\Gamma(\gamma-{\alpha\over 2})}\over
\Gamma(\gamma)}\sqrt{{(\gamma)_n(\gamma)_m}\over {n!m!}}
{}_3F_2\bigg(\matrix{-m,{\gamma-{\alpha\over 2}},{1-{\alpha\over 2}}\cr
	\gamma,1-{\alpha\over 2}-n\cr}\bigg|1\bigg).\eqno(3.11)
$$
In the case of $\alpha$ being a non-negative even number ($\alpha=2,4,6,\dots$), the Clausen hypergeometric function ${}_3F_2$ in (3.11) may be looked upon as a polynomial of degree ${\alpha\over 2}-1$ instead of an $m$-degree polynomial. This is, of course, not the case for $0<\alpha\neq 2,4,6,\dots$ in which case the numerical computational would have to be done directly using the expression (3.11). For $n\geq m$ and $\alpha=2,4,6,\dots$ we have by means of the series representation of the hypergeometric function ${}_3F_2$ that

$$
{}_3F_2\bigg(\matrix{-({\alpha\over 2}-1),\gamma-{\alpha\over
2},-m\cr
	\gamma,1-{\alpha\over 2}-n\cr}\bigg|1\bigg)=
\sum\limits_{s=0}^{{\alpha\over 2}-1}{{(-m)_s({\gamma-{\alpha\over 2}})_s(1-{\alpha\over 2})_s}\over {s!(\gamma)_s (1-{\alpha\over 2}-n)_s}}.\eqno(3.12)
$$ 
As a result, the matrix elements (3.11) further simplify into the closed form
expressions immediately appearing. These are most suitable for computational purposes as for the case of $\gamma>1$ and $\alpha = 2$, we indeed have
$$
r_{mn}^{-2}=\cases{(-1)^{m+n}
{\beta\over \gamma-1}
\sqrt{n!(\gamma)_m\over m!(\gamma)_n}& if $n> m$,\cr
\ \cr
{\beta\over \gamma-1}& if $n=m$,\cr
\ \cr
(-1)^{m+n}
{\beta\over \gamma-1}
\sqrt{m!(\gamma)_n\over n!(\gamma)_m}& if $n<m$.\cr}
\eqno(3.13)
$$
On the other hand, for $\gamma>2$ and $\alpha=4$, we have from (3.11) and (3.12) that  
$$
r_{mn}^{-4}=\cases{
{(-1)^{m+n}\beta^2\over \gamma(\gamma-1)(\gamma-2)}
\sqrt{n!(\gamma)_m\over m!(\gamma)_n}[\gamma(n-m+1)+2m]& if $n> m$,\cr
\ \cr
{\beta^2\over \gamma(\gamma-1)(\gamma-2)}[\gamma+2n]& if $n= m$,\cr
\ \cr
{(-1)^{m+n}\beta^2\over \gamma(\gamma-1)(\gamma-2)}
\sqrt{m!(\gamma)_n\over n!(\gamma)_m}[\gamma(m-n+1)+2n]& if $n<m$.\cr}
\eqno(3.14)
$$
We also point out, for $\gamma>3$ and $\alpha=6$, Eq.(3.12) lets us deduce  
$$
r_{mn}^{-6}=\cases{
{(-1)^{m+n}\beta^3\over 2(\gamma+1)\gamma(\gamma-1)(\gamma-2)(\gamma-3)}
\sqrt{{n! (\gamma)_m}\over m!(\gamma)_n}\times\cr[(2+n)(1+n)\gamma(\gamma+1)-2m(1+n)
(\gamma-3)(\gamma+1)-m(1-m)(\gamma-2)(\gamma-3)]& if $n> m$,\cr
\ \cr
{\beta^3\over (\gamma+1)\gamma(\gamma-1)(\gamma-2)(\gamma-3)}(\gamma + \gamma^2 + 6\gamma n + 6 n^2)
& if $n= m$,\cr
\ \cr
{(-1)^{m+n}\beta^3\over 2(\gamma+1)\gamma(\gamma-1)(\gamma-2)(\gamma-3)}
\sqrt{m!(\gamma)_n\over n!(\gamma)_m}\times\cr[(2+m)(1+m)\gamma(\gamma+1)-2n(1+m)(\gamma-3)(\gamma+1)-n(1-n)(\gamma-2)(\gamma-3)]& if $ n<m$.\cr}
\eqno(3.15)
$$
We can derive similar expressions for all even integers beyond 6, i.e. $\alpha=8, 10,\dots$ where we have, for $n\geq m$, that 
$$
r_{mn}^{-\alpha}=(-1)^{n+m}\beta^{{\alpha\over 2}}{{({\alpha\over 2})_n}\over
(\gamma)_n}{{\Gamma(\gamma-{\alpha\over 2})}\over
\Gamma(\gamma)}\sqrt{{(\gamma)_n(\gamma)_m}\over {n!m!}}\sum\limits_{s=0}^{{\alpha\over 2}-1}{{(-m)_s({\gamma-{\alpha\over 2}})_s(1-{\alpha\over 2})_s}\over {s!(\gamma)_s (1-{\alpha\over 2}-n)_s}},\eqno(3.16)
$$
and the matrix elements with $0\leq n<m$ are incorporated by using the symmetry property, i.e. $r^{-\alpha}_{mn}=r^{-\alpha}_{nm}$. For $N$-dimensional case, the matrix elements of the singular operator $r^{-\alpha}$ can be easily found in analogy with (3.10) 
$$
r_{mn}^{-\alpha}=(-1)^{n+m}\beta^{{\alpha\over 2}}{{({\alpha\over 2})_n}\over
(\gamma)_n}{{\Gamma(\gamma_N-{\alpha\over 2})}\over
\Gamma(\gamma_N)}\sqrt{{(\gamma_N)_n(\gamma_N)_m}\over {n!m!}}
{}_3F_2\bigg(\matrix{-m,\gamma_N-{\alpha\over
2},1-{\alpha\over 2}\cr \gamma_N,1-n-{\alpha\over 2}\cr}\bigg|1\bigg).\eqno(3.17)
$$
The results for the special cases $\alpha=2,4,6,\dots$ can be obtained in a similar fashion to 3.14-3.16 through the substitution of $\gamma$ by $\gamma_N=1+\sqrt{A+(l+{N\over 2}-1)^2}$. 
\medskip

\ni{\bf 3.2 Matrix elements of the power-law potentials $r^{q}$:}
\medskip
\noindent We now use the orthonormal eigenfunctions (3.3) to compute the matrix elements for the power-law potential operators $r^{q}$, $q=2, 4,6,\dots$. In analogy with (3.10), this can be achieve by means of the identity: for $m$ and $n$ non-negative integers and $2\gamma+q>0$ 
$$\int\limits_0^\infty r^{2\gamma+q-1} e^{-\beta r^2}
{}_1F_1\bigg(\matrix{-n\cr
	\gamma\cr}\bigg|\beta r^2\bigg){}_1F_1\bigg(\matrix{-m\cr
	\gamma\cr}\bigg|\beta r^2\bigg)dr={(-{q\over 2})_n\Gamma(\gamma+{q\over 2})\over2\beta^{\gamma+{q\over 2}}(\gamma)_n} {}_3F_2\bigg(\matrix{-m,{\gamma+{q\over 2}},{1+{q\over 2}}\cr
	\gamma,1+{q\over 2}-n\cr}\bigg|1\bigg)
\eqno(3.18)
$$
In the case of $q$ being a positive even number ($q=2,4,6,\dots$), the Clausen hypergeometric function ${}_3F_2$ in (3.18) can be further simplified. Indeed in this case, we prove the following result.
\medskip
\noindent{\bf Theorem 1:} For $t=0,1,2,\dots,{q\over 2}$ and $q=2,4,6,\dots$, the matrix elements of the power-law potential $r^{q}$, $q=2, 4,6,\dots$ in terms of the orthonormal functions (3.3) are
$$r_{mn}^q =
\cases{0,&if $n>m+{q\over 2}$\cr\cr
	{(-1)^q\over \beta^{q\over 2}}\sqrt{(\gamma+{\alpha\over 2})_m\Gamma(\gamma+{q\over 2})(m+{q\over 2})!\over (\gamma)_m\Gamma(\gamma)m!},&if $n=m+{q\over 2}$\cr\cr
	{\Gamma(\gamma+{q\over 2})(\gamma+m-t)_{q\over 2}(m-t+1)_{q\over 2}\over (-1)^{t-q}\beta^{q\over 2}t!\Gamma(\gamma) (\gamma)_{q\over 2}}
\sqrt{m!(\gamma)_m\over (\gamma)_{m+{q\over 2}-t}(m+{q\over 2}-t)!}
\sum\limits_{j=0}^{t}{(-t)_j(\gamma+m+{q\over 2}-t)_j({q\over 2}+m-t+1)_j\over (\gamma+m-t)_j(m+1-t)_j~j!},&if $n=m+{q\over 2}-t$\cr\cr
	{\Gamma(\gamma+{q\over 2})~(1-\gamma-m)_{q\over 2}~(-m)_{q\over 2}\over  (-\beta)^{q\over 2}~\Gamma(\gamma)~{q\over 2}!~(\gamma)_{q\over 2}}\sum\limits_{j=0}^{q\over 2}{(-{q\over 2})_j(\gamma+m)_j(m+1)_j\over (\gamma+m-{q\over 2})_j(m+1-{q\over 2})_j~j!},&if $m=n$\cr
\cr
{\Gamma(\gamma+{q\over 2})(\gamma+n-t)_{q\over 2}(n-t+1)_{q\over 2}\over (-1)^{t-q}\beta^{q\over 2}t!\Gamma(\gamma) (\gamma)_{q\over 2}}
\sqrt{n!(\gamma)_n\over (\gamma)_{n+{q\over 2}-t}(n+{q\over 2}-t)!}
\sum\limits_{j=0}^{t}{(-t)_j(\gamma+n+{q\over 2}-t)_j({q\over 2}+n-t+1)_j\over (\gamma+n-t)_j(n+1-t)_j~j!},&if $m=n+{q\over 2}-t$\cr\cr
{(-1)^q\over \beta^{q\over 2}}\sqrt{(\gamma+{\alpha\over 2})_n\Gamma(\gamma+{q\over 2})(n+{q\over 2})!\over (\gamma)_n\Gamma(\gamma)n!},&if $m=n+{q\over 2}$\cr\cr
0,&if $m>n+{q\over 2}$\cr
}\eqno(3.19)
$$
\noindent{\bf Proof:} From (3.18), we have
$$r_{mn}^q=(-1)^{m+n}\beta^{-{q/2}}{\Gamma(\gamma+{q\over 2})\over \Gamma(\gamma)(\gamma)_n} \sqrt{(\gamma)_n(\gamma)_m\over n!m!}\bigg(-{q\over 2}\bigg)_n~{}_3F_2\bigg(\matrix{-m,{\gamma+{q\over 2}},{1+{q\over 2}}\cr
	\gamma,1+{q\over 2}-n\cr}\bigg|1\bigg)
$$
and the problem is now reduced to the simplification of the product 
$$\bigg(-{q\over 2}\bigg)_n~{}_3F_2\bigg(\matrix{-m,{\gamma+{q\over 2}},{1+{q\over 2}}\cr
	\gamma,1+{q\over 2}-n\cr}\bigg|1\bigg).
$$
Using the series representation for ${}_3F_2$, we can write
$$I=\bigg(-{q\over 2}\bigg)_n~{}_3F_2\bigg(\matrix{-m,{\gamma+{q\over 2}},{1+{q\over 2}}\cr
	\gamma,1+{q\over 2}-n\cr}\bigg|1\bigg)=\sum\limits_{k=0}^m {(-m)_k(\gamma+{q\over 2})_k(-{q\over 2}-k)_n\over (\gamma)_k k!},$$
where we have used the identity $(a-n)_k=(1-a)_n(a)_k/(1-a-k)_n$. From the definition of the Pochhammer symbol $(-{q\over 2}-k)_n$ we have that $I=0$ for $n>m+{q\over 2}$ and for $n={q\over 2}+m$ we have
$$I={(-1)^{q\over 2}(\gamma+{q\over 2})_m(m+{q\over 2})!\over (\gamma)_m}.$$
Finally, for $n=m+{q\over 2}-t$, $t=0,1,2,\dots,{q\over 2}$, 
$$I=\sum\limits_{k=0}^m {(-m)_k(\gamma+{q\over 2})_k(-{q\over 2}-k)_{m+{q\over 2}-t}\over (\gamma)_k~k!}=\sum\limits_{k=m-t}^m {(-m)_k(\gamma+{q\over 2})_k(-{q\over 2}-k)_{m+{q\over 2}-t}\over (\gamma)_k~k!}.$$
The completion of the proof then follows by shifting the index $j=k-m+t$ of the finite sum.\qed 
\medskip
\noindent As consequence of Theorem 1, for $q=2$ and $\gamma>-1$ we have that
$$r_{mn}^2 =
\cases{0,&if $n>m+1$\cr\cr
	\beta^{-1}\sqrt{(m+1)(\gamma+m)},&if $n=m+1$\cr\cr
	\beta^{-1}(\gamma+2n),&if $n=m$\cr\cr
	\beta^{-1}\sqrt{(n+1)(\gamma+n)},&if $m=n+1$\cr\cr
	0,&if $m>n+1$.\cr}\eqno(3.20)
$$
Furthermore, the explicit formula for the matrix elements of the operator $r^{4}$ now reads 
$$r^4_{mn} =
\cases{0,&if $n>m+2$\cr\cr
       \beta^{-2}\sqrt{(m+1)(m+2)(\gamma+m)(\gamma+m+1)},&if $n=m+2$\cr\cr
	2\beta^{-2}(\gamma+2m+1)\sqrt{(m+1)(\gamma+m)},&if $n=m+1$\cr\cr
	\beta^{-2}(\gamma+6m^2+6\gamma m+\gamma^2),&if $n=m$\cr\cr
	2\beta^{-2}(\gamma+2n+1)\sqrt{(n+1)(\gamma+n)},&if $m=n+1$\cr\cr
      \beta^{-2}\sqrt{(n+1)(n+2)(\gamma+n)(\gamma+n+1)},&if $m=n+2$\cr\cr
	0,&if $m>n+1$.\cr}.\eqno(3.21)
$$
\medskip
\np 
\ni{\bf 4. Applications and Numerical Results}
\medskip 
 \medskip
\ni{\bf 4.1 Spiked Harmonic oscillator Hamiltonians}
\medskip
\noindent There are several reasons for the interest in the spiked harmonic oscillator Hamiltonian (1.2)
and its extension, the so-called generalized spiked harmonic oscillator, 
$$
H=-{d^2\over dr^2}+ Br^2+{A\over r^2}+{\lambda\over r^\alpha}.\eqno(4.1)
$$
First, it represents the simplest model of certain realistic interaction potentials in atomic, molecular and nuclear physics, and second, its interesting intrinsic properties from the viewpoint of mathematical physics: (1) an eigenvalue of the perturbed operator may not converges to the original one as $\lambda\rightarrow 0$ (the Klauder Phenomenon) and (2) the perturbation series is ordered in fractional powers of $\lambda$, and in the cases $\alpha>{5/2}$ the regular Rayleigh-Schr\"odinger perturbation theory fails badly. 

We shall consider the problem initially in $N = 3$ spatial dimensions. It was proven earlier\sref{\hale} that the set of $L_2[0,\infty)$-functions $\{\psi_n(r)\}_{n=0}^\infty$ as defined by (3.3), is a complete orthonormal basis for the Hilbert space $L_2[0,\infty)$. This basis was starting point for perturbative expansions and variational analysis of the Hamiltonians (1.2) and (4.1). The main approach of the earlier variational investigation of (1.2) was the re-writing of the Hamiltonian as
$$H\equiv -{d^2\over dr^2}+r^2+{A\over{r^2}}+\left({\lambda\over r^\alpha}-{A\over{r^2}}\right).\eqno(4.2)$$
The parameter $A$ serves as an extra degree of freedom that can be used to accelerate the convergence to the exact eigenvalues through the minimization of the eigenvalues of the diagonalizable $D\times D$ symmetric matrix.
Straightforward calculations using (3.11) and (3.13) show that the matrix elements $H_{mn}$ of the Hamiltonian (4.2) are ($m,n=0,1,2,..,D-1$, $n\geq m$)
$$\eqalign{
H_{mn}=& 2(2n+\gamma)\delta_{nm}+
(-1)^{n+m}\lambda {{({\alpha\over 2})_n}\over
(\gamma)_n}{{\Gamma(\gamma-{\alpha\over 2})}\over
\Gamma(\gamma)}\sqrt{{(\gamma)_n(\gamma)_m}\over {n!m!}}
{}_3F_2\bigg(\matrix{-m,\gamma-{\alpha\over
2},1-{\alpha\over 2}\cr
	\gamma,1-n-{\alpha\over 2}\cr}\bigg|1\bigg)\cr
&-(-1)^{m+n}
{(\gamma-{3\over 2})(\gamma-{1\over 2})\over \gamma-1}
\sqrt{n!(\gamma)_m\over m!(\gamma)_n},\cr}\eqno(4.3)
$$
where $2\gamma>\alpha$ and the matrix elements with $0\leq n<m$ are incorporated by using the symmetry property of the matrix. In order to apply the method discussed in Section 2, we write (1.2) in more extended form
$$H\equiv -{d^2\over dr^2}+Br^2+{A\over{r^2}}+(1-B)r^2+\left({\lambda\over r^\alpha}-{A\over{r^2}}\right),\eqno(4.4)$$
In this case, the matrix elements of the Hamiltonian (4.4) assume the form
$$
H_{mn}=2\beta (2n+\gamma)\delta_{mn}+(1-\beta^2)r^2_{mn}+\lambda r_{mn}^{-\alpha}-Ar_{mn}^{-2},\quad (\beta=\sqrt{B},m,n=0,1,2,\dots,D-1)\eqno(4.5)$$
where $r^2_{mn}$ is given by (3.20), $r_{mn}^{-\alpha}$ given by (3.11), and $r_{mn}^{-2}$ is given by (3.13). In order to illustrate the difference between using the expressions (4.3) and (4.5), we restrict our calculation to $\alpha=4$. The first variational approximation (subspace of dimension 1) of the ground-state eigenvalues of the spiked harmonic oscillator Hamiltonian yields by means of (4.3) the approximation
$$\epsilon_0=\min\limits_{A> 0.75}~\bigg\{\gamma+1+{\lambda\over (\gamma-1)(\gamma-2)}+{1\over 4(\gamma -1)}\bigg\},\quad (\gamma=1+{1\over 2}\sqrt{1+4A}>2)\eqno(4.6)$$
while the matrix elements (4.5) yield the approximation
$$\epsilon_0=\min\limits_{A>0.75,B>0}~\bigg\{{\gamma\over \beta}+\beta+{\lambda\beta^2\over (\gamma-1)(\gamma-2)}+{\beta\over 4(\gamma -1)}\bigg\},\quad (\gamma=1+{1\over 2}\sqrt{1+4A}>2, \beta=\sqrt{B}).\eqno(4.7)$$
The minimization of (4.6) over the parameter $A$ yields
$$\epsilon_0={5\over 2}+
{3\over 2}
\bigg(8\lambda-1+4\sqrt{4\lambda^2-\lambda}\bigg)^{1/3}+\bigg(6\lambda-{3\over 4}-3\sqrt{4\lambda^2-\lambda}\bigg)\bigg(8\lambda-1+4\sqrt{4\lambda^2-\lambda}\bigg)^{2/3}
$$
which implies the upper bound $\epsilon_0=21.427~793$ for $\lambda=1000$; while the minimization of (4.7) over $A$ and $B$  yields  $\epsilon_0=21.374~087$ with first decimal place exact. In Table I, we present a comparison between eigenvalues computation using (4.6) and (4.7) for $\lambda=0.1$ to illustrate the increased in the rate of convergence obtained when using our new approach. It should be noted that the optimization over the parameter $A$ of the $1000\times 1000$-diagonalizable matrix yields an upper bound of $E^A=3.575~557$ with $A\approx 6.076$, while the minimization and diagonalization of the symmetric matrix over the parameters $A$ and $B$ greatly reduce the number of the matrix elements needed by a ratio of approximately $10:1$. As shown in Table I, $100\times 100$ matrix is sufficient to achieve an exact eigenvalues of $E^{A,B}=3.575~552$. Because we have established simple formulas for the matrix elements in the cases $\alpha=4$ and $\alpha=6$ (given by (3.14) and (3.15) respectively), the determination of the energy values to any desired accuracy reduced to an easy task as indicated in Table II where we report our eigenvalue computation for the case $\alpha=6$ and for different values of the parameter $\lambda$. A heuristic scheme for ascertaining the eigenvalues to any required number of digits is as follows. The eigenvalues obtained from successive levels, such as ($1\times 1, 2\times 2,\dots$), of the truncated matrix are compared, and the calculation ceases when the successive eigenvalue agree with each other up to the prescribed decimal place. Further advantage of the variational approach presented here is the amount of information that we get about the spectrum of the Hamiltonian every time we compute the eigenvalues via the diagonalization and minimization. Indeed, we obtain, for $D\times D$-matrix, a set of upper bounds for the eigenvalues $E_0, E_1,\dots, E_{D-1}$. Each can be improved by either an increase in the dimension of the matrix, or by extracting the desired level through the diagonalization and subsequent minimization with respect to parameters $A$ and $B$. For the $N$-dimensional case, the matrix elements of the singular operator $r^{-\alpha}$ turn out to be
$$
\bra \psi_{ml}^N|r^{-\alpha}| \psi_{nl}^N\ket=(-1)^{n+m}\beta^{{\alpha\over 2}}{{({\alpha\over 2})_n}\over
(\gamma)_n}{{\Gamma(\gamma_N-{\alpha\over 2})}\over
\Gamma(\gamma_N)}\sqrt{{(\gamma_N)_n(\gamma_N)_m}\over {n!m!}} 
{}_3F_2\bigg(\matrix{-m,\gamma_N-{\alpha\over
2},1-{\alpha\over 2}\cr
	\gamma_N,1-{q\over 2}-n\cr}\bigg|1\bigg)
.
\eqno(4.8)$$
Matrix elements for the special cases of $\alpha=2,4,6,\dots$ are obtained  by substituting in Eqs.(3.13-15) for $\gamma$ the expression $\gamma_N$, where $\gamma_N = 1+\sqrt{A+(\Lambda+{1\over 2})^2}$. The matrix elements of the spiked harmonic oscillator Hamiltonian now turn out to be very similar to those in Eq.(4.5), namely
$$H_{mn}= 2\beta(2n+\gamma_N)\delta_{mn}+(1-B)r_{mn}^{2}+\lambda r_{mn}^{-\alpha} -Ar_{mn}^{-2}.\eqno(4.9)$$
In Table III, upper bounds  $E_{00}^N$, obtained by the optimization of the eigenvalues of a $10\times 10$-matrix over the parameters $A$ and $B$. The results are reported for the Hamiltonian
$H=-{d^2\over dr^2}+{\Lambda(\Lambda+1)\over r^2}+r^2+{1000\over r^{4}}$ where $\Lambda={1\over 2}(N+2l-3)$ for dimension $N=2$ to $10$ with the angular momentum $l=0$.
\medskip
\ni{\bf 4.2 Anharmonic Singular Hamiltonian}
\medskip
\noindent The higher-order anharmonic singular Hamiltonians (1.3) have attracted much attention recently\sref{\za-\gotc}. This is in part because the study of the relevant Schr\"odinger equation with anharmonic potentials provides understanding and insight for the corresponding physical problems, and also because the determination of its energy is itself a challenging problem. In 3-dimensional space, there are two main methods for dealing with the anharmonic potentials $V(r)=ar^2+b r^{-4}+cr^{-6}.$
A method due to Varshni\sref{\va} is based on an {\it ansatz} for the eigenfunctions, sufficient conditions on parameters to yield exact solutions, and a limit from initial box confinement.  The other method, mainly proposed by Znojil\sref{\za-\zc}, relies on a Laurent series {\it ansatz} for the eigenfunctions, which converts the Schr\"odinger equation into a difference equation which is solved by the use of continued fractions. An interesting study related to Varshni's idea\sref{\va} for the potential $V(r)$ in 2-dimensions was proposed recently by Shi-Hai Dong and Zhong-Qi Ma\sref{\sz}. 

The method discussed in Section 2 of the present article provides a uniformly simple, straightforward and very efficient way of yielding accurate energies of the entire spectrum of the anharmonic potentials $V(r)$ not only in 1 or 2 dimensions but actually in arbitrary dimensions with arbitrary angular momentum number $l=0,1,2,\dots$. We start with a wider class of anharmonic singular Hamiltonian given by
$$H=-{d^2\over dr^2}+{\Lambda(\Lambda+1)\over r^2}+a_1r^2+{a_2\over r^2}+{a_3\over r^4}+{a_4\over r^6}\eqno(4.10)$$
where $\Lambda={1\over 2}(M-3)$, and $M=N+2l$. Clearly, the case $V(r)=a_1r^2+a_3 r^{-4}+a_4r^{-6}$ appears as special case with $a_2=0$. Following the procedure discussed in Section 2, we write the Hamiltonian as
$$H=-{d^2\over dr^2}+Br^2+{\Lambda(\Lambda+1)+A\over r^2}+(a_1-B)r^2+{(a_2-A)\over r^2}+{a_3\over r^4}+{a_4\over r^6}.\eqno(4.11)$$
The matrix elements of the Hamiltonian (4.11) take the form ($m,n=0,1,2,\dots$)
$$H_{mn}=2\sqrt{B} (2n+\gamma_N)\delta_{mn}+(a_1-B)r_{mn}^2+(a_2-A)r_{mn}^{-2}+a_3r_{mn}^{-4}+a_4r_{mn}^{-6}\eqno(4.12)$$
for $\gamma>3$. To analyze the precision of the method proposed here, we compare our results with some special cases for which the exact eigenvalues are known. The case of $a_1=a_3=a_4=1,\ a_2=0,$ which yields the ground-state energy $E=5$ has been analyzed by Znojil\sref{\zd}, Guardiola and Ros\sref{\gr}, and Buend\'ia {\it et al}\sref{\bgp} by different techniques. Table (IV) shows the exact eigenvalues of $E=5$ can be reached with the diagonalization of $50\times 50$-matrix. It should be noted however we have fixed the dimension of the matrix to $50\times 50$, but the particular value $E=5$ can be reached with far fewer matrix elements, indeed a $30\times 30$ matrix is sufficient to achieve such accuracy. Further, the exact energies of $7,7,11$ corresponding to $(a_1,a_3,a_4)=(1,9,9), (1,-7,49)$, and $(1,45, 225)$ respectively follow simply with the optimization of the diagonalizable $40\times 40$-matrix with $(A,B)=(17.47,5.69), (18.86,5.53)$, and $(17.92,5.40)$.  These results simply indicate the generality and the efficient of our approach. Note, the case of $(a_1,a_3,a_4)=(1,-7,49)$ also reflect the applicability of the method in the case of the parameter $a_3$ is negative. It is quite clear from Tables (IV) and (V) the generality of the method proposed here. In Table (V), we illustrate the applicability of the method to the problem of obtaining the energies in different dimensions. Similar results for different excited states can be easily reproduce. All the eigenvalues quoted in Tables I-V agree with the numerical solutions of the correspondence Schr\"odinger equation. Generally speaking, the precision of the energies to any number of decimal places can be easily achieve by increasing the dimension of the matrix. 
\medskip
\ni{\bf 5. Conclusion}
\medskip
\noindent 
We have developed an effective variational method to study a large family of singular Hamiltonians. A key feature of this work is the establishment and simplification of closed-form analytical expressions for the matrix elements with respect to a basis derived from a soluble singular problem.  These formulas are general in the sense that they include two pameters from the basis which can then be used to optimize the matrix eigenvalues obtained for the problem in hand. The improved variational approach yields faster energy convergence than was possible earlier.
\bigskip
\noindent {\bf Acknowledgment}
\medskip
\noindent Partial financial support of this work under Grant Nos. GP3438 and GP249507 from the 
Natural Sciences and Engineering Research Council of Canada is gratefully 
acknowledged by two of us (respectively [RLH] and [NS]).
\vfil\eject
\noindent{\tiny {\bf Table (I):}~~~Upper bounds $E^A$ for 
$H=-\Delta+r^2+{1\over 10 r^{4}}$ are obtained by diagonalization then minimization of the $D\times D$ matrix, only over the parameter $A$.  $E^{A,B}$ are the corresponding values minimized over both $A$ and $B.$   The eigenvalue $3.575~552$ (exact to 7 places) can be easily verified by direct numerical integration of Schr\"odinger's equation. } 
\bigskip 

\hskip 1 true in
\vbox{\tabskip=0pt\offinterlineskip
\def\tablerule{\noalign{\hrule}}
\def\vr{\vrule height 12pt}
\halign to300pt{\strut#\vr&#
\tabskip=1em plus2em
&\hfil#\hfil
&\vrule#
&\hfil#\hfil
&\vrule#
&\hfil#\hfil
&\vr#\tabskip=0pt\cr
\tablerule&&$D\times D$&&\bf $E^A$&&\bf $E^{A,B}$ &\cr\tablerule
&&$1\times 1$&&$3.745~811$&&$3.664~281$&\cr
&&$~$&&$(A\approx 1.52)$&&$(A\approx1.92,B\approx 1.62)$&\cr\tablerule
&&$10\times 10$&&$3.602~189$&&$3.582~194$&\cr
&&$~$&&$(A\approx 2.84)$&&$(A\approx 4.75,B\approx 11.66)$&\cr\tablerule
&&$20\times 20$&&$3.588~143$&&$3.576~773$&\cr
&&$~$&&$(A\approx 3.68)$&&$(A\approx 8.14,B\approx 32.22)$&\cr\tablerule
&&$100\times 100$&&$3.577~007$&&$3.575~552$&\cr
&&$~$&&$(A\approx 7.44)$&&$(A\approx 9.73,B\approx 297.23)$&\cr\tablerule
&&$200\times 200$&&$3.576~015$&&$3.575~552$&\cr
&&$~$&&$(A\approx 10.39)$&&$(A\approx 3.76,B\approx 873.58)$&\cr\tablerule
}}
\vfil\eject
\noindent{\tiny {\bf Table (II):}~~~Upper bounds $E^A$ for 
$H=-\Delta+r^2+{\lambda\over r^{6}}$ obtained by diagonalization followed by minimization of the $D\times D$ matrix, only over the parameter $A$. $E^{A,B}$ are the corresponding values minimized over both $A$ and $B.$  The results are displayed for different values of the potential coupling $\lambda$. The eigenvalues are correct for the 7 digits, as can be easily verify by direct numerical integration of Schr\"odinger's equation.} 
\bigskip 

\hskip 0.5 true in
\vbox{\tabskip=0pt\offinterlineskip
\def\tablerule{\noalign{\hrule}}
\def\vr{\vrule height 12pt}
\halign to350pt{\strut#\vr&#
\tabskip=1em plus2em
&\hfil#\hfil
&\vrule#
&\hfil#\hfil
&\vrule#
&\hfil#\hfil
&\vr#\tabskip=0pt\cr
\tablerule&&$\lambda$&&\bf $E^A$&&\bf $E^{A,B}$ &\cr\tablerule
&&$1000$&&$12.718~617$&&$12.718~617$&\cr
&&$~$&&$(32\times 32,\ A\approx 20.52)$&&$(15\times 15,\ A\approx54.41,B\approx4.51)$&\cr\tablerule
&&$100$&&$8.413~358$&&$8.413~358$&\cr
&&$~$&&$(65\times 65,\ A\approx 19.61)$&&$(22\times 22,\ A\approx 8.88,B\approx 9.76)$&\cr\tablerule
&&$10$&&$6.003~209$&&$6.003~209$&\cr
&&$~$&&$(150\times 150,\ A\approx 9.71)$&&$(30\times 30,\ A\approx 9.66,B\approx 21.41)$&\cr\tablerule
&&$1$&&$4.659~940$&&$4.659~940$&\cr
&&$~$&&$(350\times 350,\ A\approx 17.79)$&&$(45\times 45,\ A\approx 18.34,B\approx  56.77)$&\cr\tablerule
&&$0.1$&&$3.915~665$&&$3.915~665$&\cr
&&$~$&&$(1000\times 1000)$&&$(80\times 80,\ A\approx  4.47,B\approx 176.63)$&\cr\tablerule
&&$0.01$&&$3.505~492$&&$3.505~455$&\cr
&&$~$&&$(1000\times 1000)$&&$(100\times 100,\ A\approx 20.88,B\approx 348.92)$&\cr\tablerule
}}
\vfil\eject

\vfil\eject
\noindent{\tiny {\bf Table (III)}~~~Upper bounds $E_{00}^N$ for 
$H=-{d^2\over dr^2}+{\Lambda(\Lambda+1)\over r^2}+r^2+{1000\over r^{4}}$ for dimension $N=2$ to $10$, 
obtained by diagonalization then minimization of the $10\times 10$ matrix} over $A$ and $B$.\bigskip 
\hskip 1 true in
\vbox{\tabskip=0pt\offinterlineskip
\def\tablerule{\noalign{\hrule}}
\def\vr{\vrule height 12pt}
\halign to300pt{\strut#\vr&#
\tabskip=1em plus2em
&\hfil#\hfil
&\vrule#
&\hfil#\hfil
&\vr#\tabskip=0pt\cr
\tablerule&&$N$&&\bf $E_{00}^N$&\cr\tablerule
&&2&&$21.350~246\ (A\approx 71.44, B\approx 1.775)$&\cr\tablerule
&&3&&$21.369~463\ (A\approx 71.27, B\approx 1.774)$&\cr\tablerule
&&4&&$21.427~056\ (A\approx 70.79, B\approx 1.772)$&\cr\tablerule
&&5&&$21.522~860\ (A\approx 69.96, B\approx 1.769)$&\cr\tablerule
&&6&&$21.656~596\ (A\approx 68.82, B\approx 1.764)$&\cr\tablerule
&&7&&$21.827~883\ (A\approx 67.34, B\approx 1.757)$&\cr\tablerule
&&8&&$22.036~232\ (A\approx 65.45, B\approx 1.749)$&\cr\tablerule
&&9&&$22.281~057\ (A\approx 63.35, B\approx 1.740)$&\cr\tablerule
&&10&&$22.561~680\ (A\approx 60.81, B\approx 1.726)$&\cr\tablerule
}}
\bigskip
\vfil\eject
\noindent{\tiny {\bf Table (IV):}~~~Upper bounds for the Hamiltonian $V(r)=ar^2+br^{-4}+cr^{-6}$ for different values of the parameters $a,b,$ and $c$. The eigenvalues are exact for the 7 digits shown, as confirmed numerically (or, for the first row, known exactly).} 

\bigskip

\hskip 0.2 true in
\vbox{\tabskip=0pt\offinterlineskip
\def\tablerule{\noalign{\hrule}}
\def\vr{\vrule height 12pt}
\halign to400pt{\strut#\vr&#
\tabskip=1em plus2em
&\hfil#\hfil
&\vrule#
&\hfil#\hfil
&\vrule#
&\hfil#\hfil
&\vrule#
&\hfil#\hfil
&\vr#\tabskip=0pt\cr
\tablerule&&$a$&&$b$&&$c$&&$E^U$&\cr\tablerule
&&1&&$1$&&$1$&&$5.000~000$&\cr
&&~&&$~$&&$~$&&$(50\times 50,\ A\approx 25.51,\ B\approx 54.35)$&\cr\tablerule
&&1&&$10$&&$1$&&$6.679~054$&\cr
&&~&&$~$&&$~$&&$(50\times 50,\ A\approx 9.65,\ B\approx 32.24)$&\cr\tablerule
&&1&&$1$&&$10$&&$6.140~123$&\cr
&&~&&$~$&&$~$&&$(50\times 50,\ A\approx  5.56,\ B\approx 38.67)$&\cr\tablerule
&&1&&$10$&&$10$&&$7.138~261$&\cr
&&~&&$~$&&$~$&&$(50\times 50,\ A\approx 7.83,\ B\approx  31.94)$&\cr\tablerule
&&1&&$100$&&$100$&&$11.791~771$&\cr
&&~&&$~$&&$~$&&$(50\times 50,\ A\approx 5.15,\ B\approx 5.00)$&\cr\tablerule
&&1&&$1000$&&$1000$&&$21.885~192$&\cr
&&~&&$~$&&$~$&&$(50\times 50,\ A\approx 19.42,\ B\approx 5.00)$&\cr\tablerule}}

\vfil\eject
\vfil\eject
\noindent{\tiny {\bf Table (V):}~~~Upper bounds $E^N$ for 
$H=-{d^2\over dr^2}+{\Lambda(\Lambda+1)\over r^2}+r^2+{1\over r^4}+{1000\over r^{6}}$ for dimension $N=2$ to $10$, 
obtained by diagonalization then minimization with respect $A$ and $B$ of a $30\times 30$ matrix. }
\bigskip 
\hskip 1 true in
\vbox{\tabskip=0pt\offinterlineskip
\def\tablerule{\noalign{\hrule}}
\def\vr{\vrule height 12pt}
\halign to300pt{\strut#\vr&#
\tabskip=1em plus2em
&\hfil#\hfil
&\vrule#
&\hfil#\hfil
&\vr#\tabskip=0pt\cr
\tablerule&&$N$&&\bf $E^N$&\cr\tablerule
&&2&&$12.704~404\ (A\approx 5.05, B\approx 5.53)$&\cr\tablerule
&&3&&$12.735~264\ (A\approx 11.73, B\approx 6.40)$&\cr\tablerule
&&4&&$12.827~666\ (A\approx 5.05, B\approx 7.82)$&\cr\tablerule
&&5&&$12.981~081\ (A\approx  5.97, B\approx 7.26)$&\cr\tablerule
&&6&&$13.194~635\ (A\approx 6.25, B\approx 7.28)$&\cr\tablerule
&&7&&$13.467~115\ (A\approx 6.12, B\approx 7.35)$&\cr\tablerule
&&8&&$13.796~990\ (A\approx 6.16, B\approx 7.46)$&\cr\tablerule
&&9&&$14.182~423\ (A\approx  6.31, B\approx 7.32)$&\cr\tablerule
&&10&&$14.621~300\ (A\approx 3.11, B\approx 7.23)$&\cr\tablerule
}}
\vfil\eject
\references{1}

\end